\documentstyle[12pt]{article}
\begin{document}
\title{\large \hspace{10cm} ITEP-15/99 \\ \hspace{10cm} April 1999 \\
\vspace{1cm}
\LARGE \bf
A remark on collisions of domain walls in a supersymmetric model}
\author {V. A. GANI\thanks{E-mail: gani@heron.itep.ru}{\,}
\\
{\it Moscow State Engineering Physics Institute (Technical University),}\\
{\it Kashirskoe shosse, 31, Moscow, 115409, Russia}\\
{\it and}\\
{\it Institute of Theoretical and Experimental Physics, Russia}\\
\\
A. E. KUDRYAVTSEV \thanks{E-mail: kudryavtsev@vitep5.itep.ru}
\\
{\it Institute of Theoretical and Experimental Physics,}\\
{\it B.Cheremushkinskaja, 25, Moscow, 117259, Russia}\\
}
\date{}
\maketitle
\vspace{1mm}
\centerline{\bf {Abstract}}
\vspace{3mm}

     The process of collision of two parallel domain walls in a
supersymmetric model is studied both in effective Lagrangian approximation
and by numerical solving of the exact classical field problem. For small
initial velocities we find that the walls interaction looks like elastic
reflection with some delay. It is also shown that in such approximation
internal parameter of the wall may be considered as a time-dependent
dynamical variable.

\newpage

     Now it is well-known that in a wide class of supersymmetric theories
families of the so-called BPS domain walls exist, see e.g.~\cite{SV,V,S}
and references therein.
These families of domain walls link various supersymmetric vacua.
Although the internal structure of each domain wall within the family is
different, they
all have the same energy. So each family of domain walls is degenerate
with respect to at least one parameter transformation, which may be
considered as a label of each specific domain wall configuration.
To be more concrete let us consider theory described by the superpotential
$$
W(\Phi,X)=\frac{m^2}{\lambda}\Phi-\frac{1}{3}\lambda\Phi^3-\alpha\Phi X^2 \ ,
\eqno(1)
$$
where $m$ is a mass parameter and $\alpha$ and $\lambda$ are coupling
constants. We assume that $\alpha$ and $\lambda$ are real and positive.
The Lagrangian for the real parts of the scalar fields is given for this
theory by the expression
$$
L=(\partial\phi)^2+(\partial\chi)^2
-\left(\frac{m^2}{\lambda}-\lambda\phi^2-\alpha\chi^2\right)^2
-4\alpha^2\phi^2\chi^2 \ .
\eqno(2)
$$
The potential term of Eq.~(2) has four degenerate vacuum states, shown in
Fig.~1. This theory posses a wide class of domain walls, which link different
vacua of the theory. These domain wall configurations may be obtained as
solutions of Bogomol'nyi - Prasad - Sommerfeld (BPS, Ref.~\cite{BPS})
equations, which look in the case of Eq.~(2) like
$$
\begin{tabular}{l}
$\displaystyle\frac{df}{dz}=1-f^2-h^2$ ,\\
$\displaystyle\frac{dh}{dz}=-\frac{2}{\rho}fh \ .
\phantom{\frac{A^{A^{A^A}}}{A}}$\\
\end{tabular}
\eqno(3)
$$
Here $\rho=\lambda/\alpha$, $m=1$ and $z$ is a space coordinate.
As it was shown in Refs.~\cite{SV,TV}, for the case $\rho=4$ solutions for
$f(z)$ and $h(z)$ may be obtained in analytical form:
$$
f(z)=\frac{a(e^{2z}-1)}{a+2e^z+ae^{2z}} \ ,
\ \ h^2(z)=\frac{2e^z}{a+2e^z+ae^{2z}} \ ,
\eqno(4)
$$
where $a$ is continuous parameter, $0\le a\le+\infty$.

     The solution (4) links vacua states 1 and 2. Depending on $a$
the specific form of $f(z)$ and $h(z)$ looks quite different. Considering
region $0<a<1$, we may introduce different parameterization:
$$
\cosh{s}=\frac{1}{a} \ .
\eqno(5)
$$
In terms of the parameter $s$ the functions $f(z)$ and $h(z)$ look like
$$
f(z)=\frac{1}{2}\left(\tanh\frac{z-s}{2}+\tanh\frac{z+s}{2}\right) \ ,
$$
$$
h^2(z)=\frac{1}{2}\left(1-\tanh\frac{z-s}{2}\tanh\frac{z+s}{2}\right) \ .
\eqno(6)
$$
From Eqs.~(6) it is clear that at large $s$ functions $f$ and $h$ split into
two orthogonal to z-axis elementary walls corresponding to the transitions
$1\to 3$ at $z=-s$ and
$3\to 2$ at $z=s$. Thus at large $s$ Eqs.~(6) describe two far
separated domain walls. The purpose of this research is to study the
dynamics of the collision between these two separated domain walls $1\to 3$
and $3\to 2$. Evidently, this question is outside of the BPS approximation
and to study this problem we should solve exact field equations, which
follow form the Lagrangian (2).

     Recently in Ref.~\cite{TV} the problem of intersection of two domain
walls in this model was considered. The authors didn't solve the exact
equations of motion, but used more simple reasonable approach considering
the parameter $a$ as dynamical effective variable. In what follows we shall
use both methods, considering $a$ (or $s$) as a function of time $t$, as
well as solving exact equations of motion for fields $f$ and $h$.
We shall demonstrate that effective Lagrangian method is consistent with
solution of Cauchy problem for exact field system.

     In terms of $a$ (5) the energy of the domain wall configuration (4)
has the form
$$
E=E_0+E_1 \ , \quad E_1=\frac{1}{2}\int\limits_{-\infty}^{+\infty}\dot{a}^2
\frac{a\cosh^3{z}+\cosh{2z}}{(a\cosh{z}+1)^4}dz \ .
\eqno(7)
$$
Here $E_0$ does not depend on $a$ and is sum of energy densities of two far
separated walls $1\to 3$ and $3\to 2$. Its specific form is inessential for
our future consideration.
Replacing $a(t)$ by $s(t)$ according to Eq.~(5), we get the following
effective Lagrangian for the new dynamical variable $s$:
$$
L_{eff}=\frac{1}{2}m(s)\dot{s}^2 \ ,
\eqno(8)
$$
where
$$
m(s)=2\left[s\tanh{s}+\frac{5}{3}-\frac{2s}{\tanh{s}}+\frac{1}{\tanh^2{s}}
\left(\frac{s}{\tanh{s}}-1\right)\right] .
\eqno(9)
$$
The effective Lagrangian (8) yields the following differential equation for
the function $s(t)$:
$$
m(s)\ddot{s}+\frac{1}{2}m^{\prime}(s)\dot{s}^2=0 \ .
\eqno(10)
$$
To observe the process of the walls collision we have to start with initial
configuration (4) with some $s(0)\gg 1$ and $\dot{s}(0)<0$. As the parameter
$s$ has a meaning of distance between the walls, such configuration
corresponds to $1\to 3$ and $3\to 2$ walls moving along $z$-axis towards
each other. Numerical solving of Eq.~(10) shows that $s(t)$ decreases to zero.
In the range of small $s$ the solution can be obtained analytically:
$$
m(s)\approx\frac{16}{15}s^2 \ , \quad
s(t)=\sqrt{2s(t_*)\dot{s}(t_*)(t-t_*)+s^2(t_*)} \ ,
\eqno(11)
$$
where $t_*$ denotes the moment of transition from numerical solving of
Eq.~(10) to analytical solution (11). While $s$ decreases from $+\infty$
to 0, the parameter $a$ changes from 0 to 1. At $s=0$
(point $A$ in Fig.~2) we have for the fields:
$$
f(z)=\tanh(z/2) \ , \quad h(z)=\frac{1}{\sqrt{2}\cosh{(z/2)}} \ .
\eqno(12)
$$
In the range $1<a<+\infty$ the parameter $s$, defined by Eq.~(5), becomes
pure imaginary. Therefore it is suitable to introduce $\tilde{s}$ as
$$
\cos{\tilde{s}}=\frac{1}{a} \ .
\eqno(13)
$$
The effective Lagrangian for $\tilde{s}$ is analogous to (8):
$$
\tilde{L}_{eff}=\frac{1}{2}\tilde{m}(\tilde{s})\dot{\tilde{s}}^2
\eqno(14)
$$
with
$$
\tilde{m}(\tilde{s})=2\left[\tilde{s}\tan{\tilde{s}}-\frac{5}{3}
+\frac{2\tilde{s}}{\tan{\tilde{s}}}+\frac{1}{\tan^2{\tilde{s}}}
\left(\frac{\tilde{s}}{\tan{\tilde{s}}}-1\right)\right] .
\eqno(15)
$$
At the moment when $a=1$ (or $s=0$) we pass from $s$ to $\tilde{s}$ which
increases from 0 to $\pi/2$. At $\tilde{s}=\pi/2$ (point $B$ in Fig.~2)
the fields $f$ and $h$ have the form:
$$
f(z)=\tanh{z} \ , \quad h(z)\equiv 0 \ .
\eqno(16)
$$
After that, $\tilde{s}$ decreases to 0 and becomes pure imaginary.
Therefore at the moment of $\tilde{s}=0$ (point $C$ in Fig.~2)
we return back to $s$, which begins to increase from 0 to $+\infty$.
Note, that for $\tilde{s}$ we also used analytical solutions
at $\tilde{s}\ll 1$ and $|\tilde{s}-\pi/2|\ll 1$. The summary
$(s,\tilde{s})$ $t$-dependencies for two initial relative velocities
$\dot{s}(0)=0.05$ and $\dot{s}(0)=0.1$ are represented in Fig.~3.

     Developed dynamical variable approximation seems to be reasonable,
but needs in numerical confirmation. We performed calculation of evolution of
the described initial configuration by solving numerically the Cauchy problem
for the system of two differential equations in partial derivatives for
the fields $f$ and $h$. These equations are consequence of the Lagrangian
(2). Results of the numerical solving of the Cauchy problem for fields
$f$ and $h$ were compared with the fields profiles obtained for current
value of the parameter $(s,\tilde{s})$ at all times $t$. For not too large
values of $\dot{s}(0)$ we observed good agreement.
It confirms that the parameter $(s,\tilde{s})$ is a good dynamical variable
for describing the process of the BPS domain walls collision in the SUSY
model that was considered.

     Note, that with $\dot{s}(0)$ being increased in numerical calculations
of the exact field problem we observe progressing difference from the field
profiles restored with the help of the dynamical variable method.
The question whether it is a problem of computational scheme or it is related
with excitation of other than zero mode field degrees of freedom is under
consideration now.

\begin{center}
\bf
Acknowledgments
\end{center}

     We are thankful to M.~B.~Voloshin for useful discussions. We would
also like to thank A.~A.~Panfilov for his help with computer graphics.

     This work was supported in part by the Russian Foundation for Basic
Research under grants No~98-02-17316 and No~96-15-96578 (both authors).
The work of V.~A.~Gani was also supported by the INTAS Grant No~96-0457
within the research program of the International Center for Fundamental
Physics in Moscow. One of the authors (A.~E.~Kudryavtsev) would like to
thank RS Grant for financial support.

\newpage

\begin{center}
\bf
Figure captions
\end{center}
\bigskip

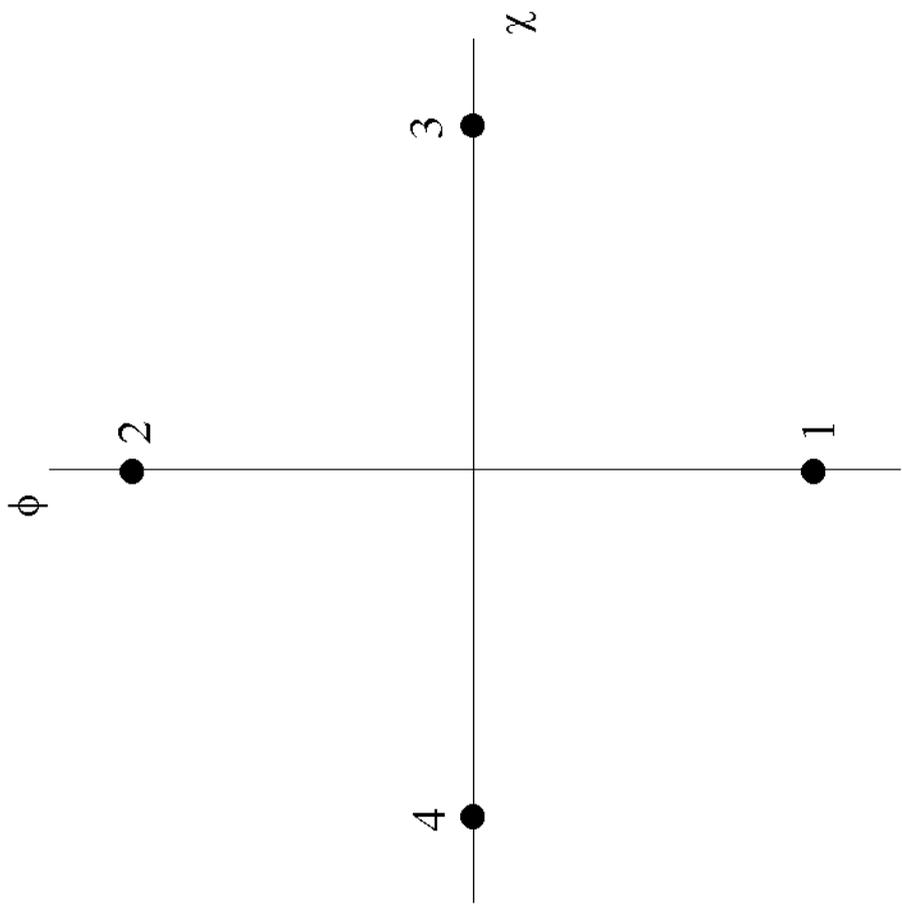
{\bf Fig.~1.} Locations of the vacuum states of the model.

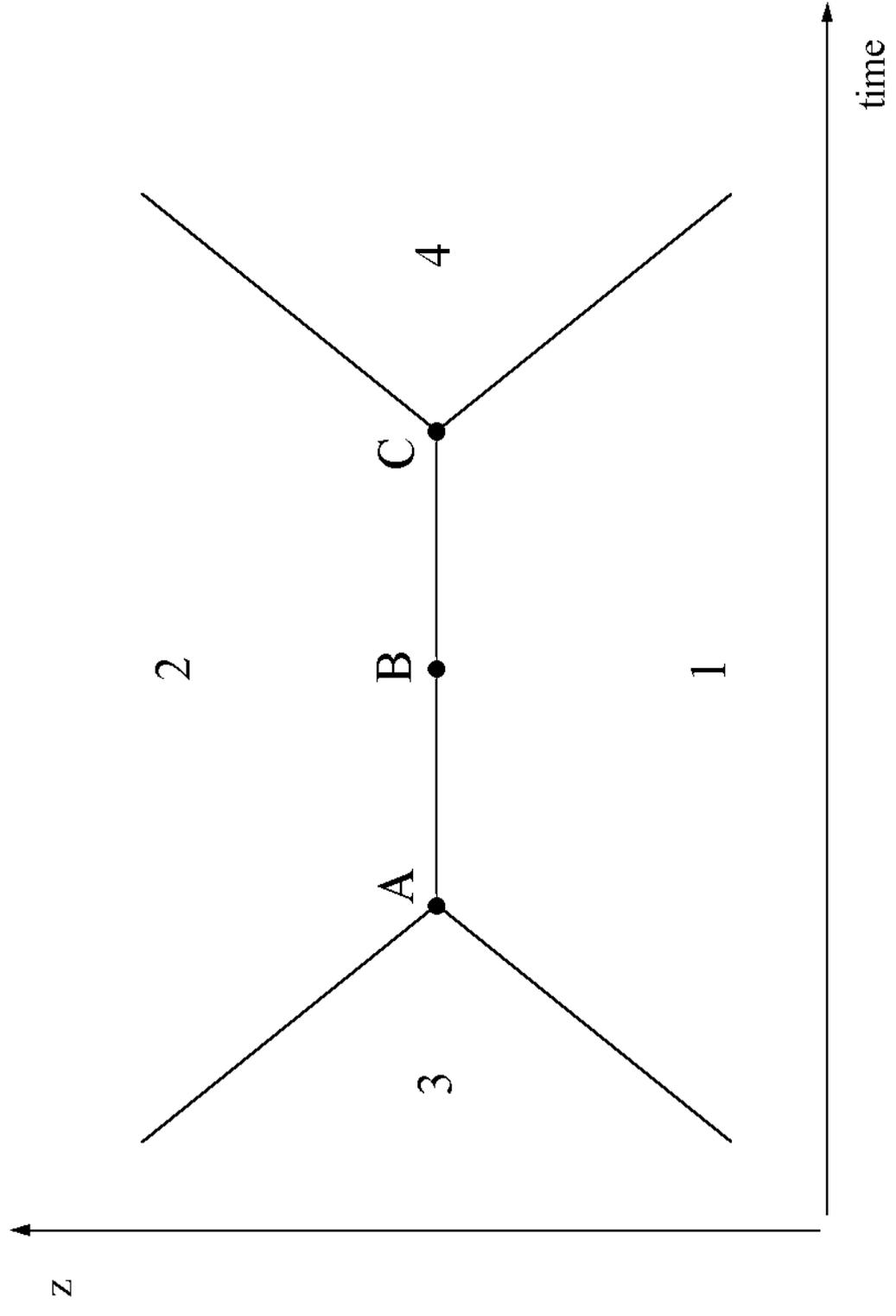
{\bf Fig.~2.} A scheme of the $1\to 3$ and $3\to 2$ walls collision process.
              Numbers 1, 2, 3 denote corresponding vacua.

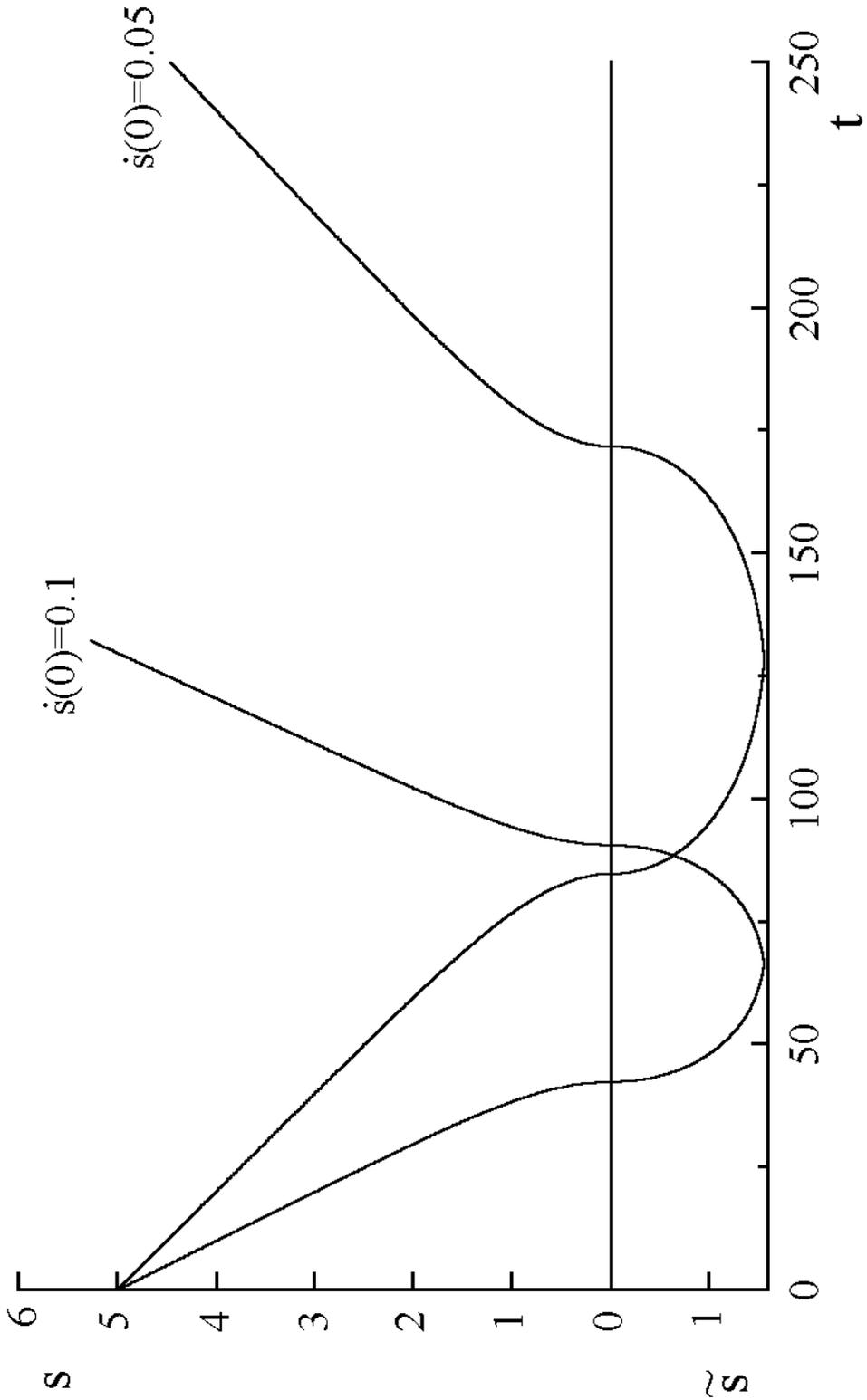
{\bf Fig.~3.} The dynamical variable ($s,\tilde{s}$) as a function of
              time for two different values of the initial velocity
              $\dot{s}(0)$.

\end{document}